\documentclass[%
reprint,
amsmath,amssymb,
aps,
]{revtex4-2}

\usepackage[english]{babel}

\usepackage{graphicx}
\usepackage{dcolumn}
\usepackage{bm}
\usepackage{amsmath}
\usepackage{cases}
\usepackage{color}

\usepackage{xr}

\makeatletter
\newcommand*{\addFileDependency}[1]{
  \typeout{(#1)}
  \@addtofilelist{#1}
  \IfFileExists{#1}{}{\typeout{No file #1.}}
}
\makeatother

\newcommand*{\myexternaldocument}[1]{%
    \externaldocument{#1}%
    \addFileDependency{#1.tex}%
    \addFileDependency{#1.aux}%
}

\myexternaldocument{Supplementary}

\begin{document}


\title{Gliders in shape-changing active matter}

\author{Akash Vardhan}
\affiliation{%
 School of Physics, Georgia Institute of Technology, Atlanta, Georgia 30332, USA}%
\author{Ram Avinery}%
\affiliation{%
 School of Physics, Georgia Institute of Technology, Atlanta, Georgia 30332, USA}%
 \author{Hridesh Kedia}%
\affiliation{%
 School of Physics, Georgia Institute of Technology, Atlanta, Georgia 30332, USA}%
 \author{Shengkai Li}%
\affiliation{%
 School of Physics, Georgia Institute of Technology, Atlanta, Georgia 30332, USA}%
 
\author{Kurt Wiesenfeld}
\affiliation{%
 School of Physics, Georgia Institute of Technology, Atlanta, Georgia 30332, USA}%
 
 \author{Daniel I. Goldman}%
\affiliation{%
 School of Physics, Georgia Institute of Technology, Atlanta, Georgia 30332, USA}%

\date{\today}

\begin{abstract}

We report in experiment and simulation the spontaneous formation of dynamically bound pairs of shape changing smarticle robots undergoing locally repulsive collisions. Borrowing terminology from Conway's simulated Game of Life, these physical `gliders' robustly emerge from an ensemble of individually undulating three-link two-motor smarticles and can remain bound for hundreds of undulations and travel for multiple robot dimensions. Gliders occur in two distinct binding symmetries and form over a wide range of angular flapping extent. This parameter sets the maximal concavity which influences formation probability and translation characteristics. Analysis of dynamics in simulation reveals the mechanism of  effective dynamical attraction -- a result of the emergent interplay of appropriately oriented and timed repulsive interactions.

\end{abstract}

\maketitle

Active matter studies focusing on collisional interactions among entities reveal fascinating emergent behaviors. Molecular motors \cite{schaller2010polar,sumino2012large}, self-propelled rods, \cite{bar2019self} and shaken grains \cite{deseigne2010collective,kudrolli2008swarming} demonstrate that patterns can form and overall order can emerge from purely repulsive interactions. And active extended objects undergoing shape-changing motions have appeared as a novel class of systems with their own repertoire of rich behavior,especially in regard to their locomotion \cite{goldman2014colloquium, lauga2008no, marvi2014sidewinding, wagner2013crawling, savoie2018phototactic}. Novel behaviors such as  mechanical diffraction \cite{schiebel2019mechanical, rieser2019dynamics,zhang2021friction} and emergent locomotion \cite{warkentin2018locomoting, savoie2019robot, brandenbourger2021active}  have been observed in shape-changing active matter systems. Furthermore, emergent collective behaviors such as collision-mediated gait synchronization in clusters of nematode worms \cite{yuan2014gait} and in undulatory robots \cite{zhou2021collective}, and collective transport by worm blobs \cite{ozkan2021collective},  have also been observed in this novel class of systems.

\begin{figure}[h!]
  \centering
  \includegraphics[width=1.0\columnwidth]{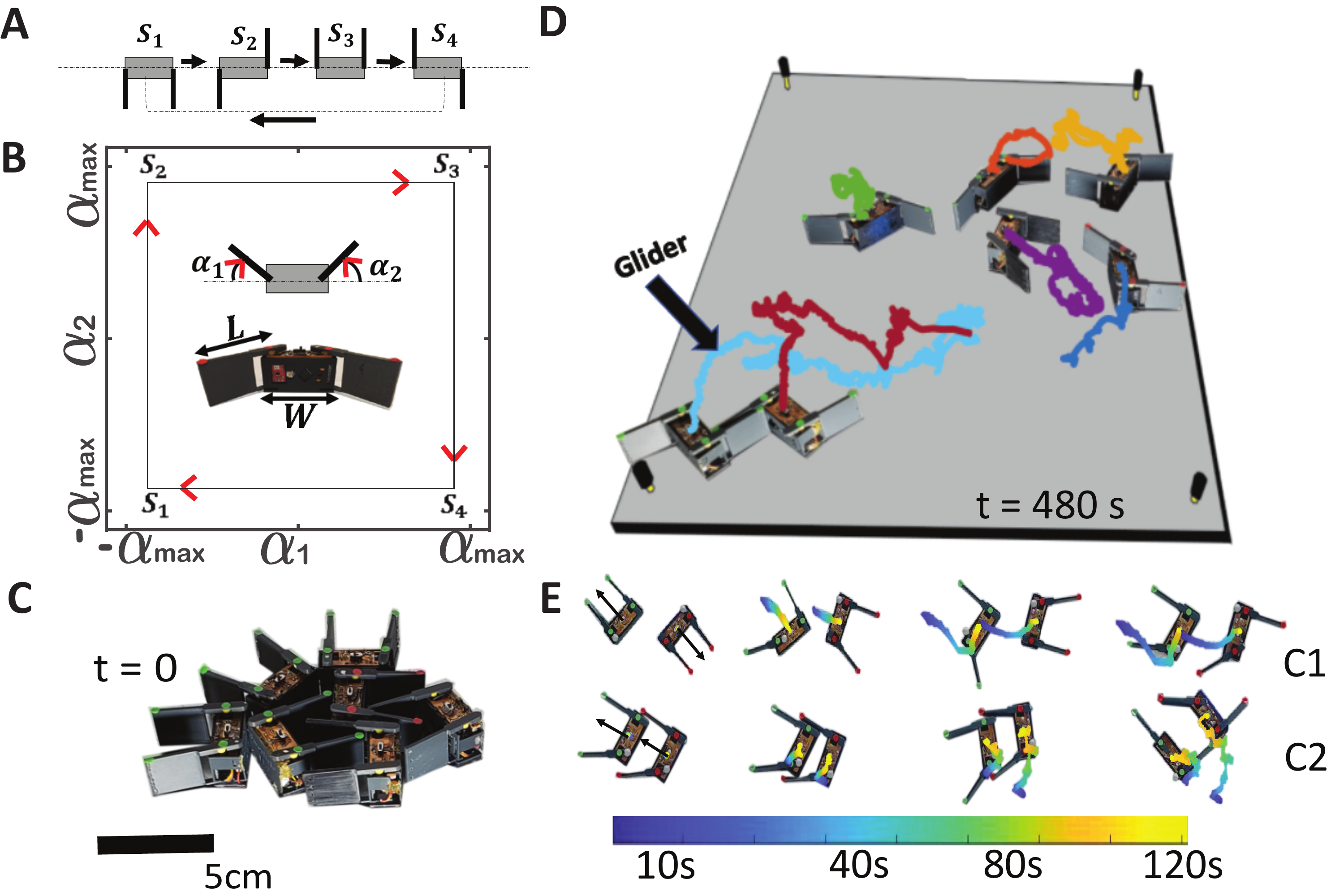}
  \caption{ \textbf{Observation of gliders in relaxation experiments.} (A) Configurations of a single smarticle at the corners of a square gait.
  (B) Each smarticle performs a continuous square gait in its shape space ($\alpha_1, \alpha_2$) with amplitude $ \alpha_{max} $. Arm angles are defined such that equal values are mirror-symmetric configurations. A smarticle rests on its middle body with its arms mounted slightly raised above the ground. The direction of red arrows represent the chirality of the gait executed.
  (C) An unconfined collective of 7 smarticles in a compressed configuration, at the beginning of the experiment ($t=0$). 
  (D) Spontaneous emergence of a locomoting bound pair (glider) from the collective as seen from smarticle body center tracks.
  (E) Snapshots of exemplary glider configurations C1 and C2. Arrows represent the body normal vectors of individual smarticles; colorbar is time in seconds. }
    \label{fig:Discovery}
\end{figure}
{Smarticles}---smart active particles \cite{savoie2015smarticles, savoie2018phototactic, warkentin2018locomoting, savoie2019robot, chvykov2021low,ozkan2021collective}, are a convenient system to study shape-changing active matter.  These two motor, 3 link robots resemble Purcell's swimmer \cite{purcell1977life} in their structure and their gait. However, they differ from swimmers at low Reynolds number \cite{purcell1977life, lauga2008no} or their terrestrial counterparts \cite{wagner2013crawling}, since an isolated smarticle has limited motility as its arms do not interact with flat ground (see Fig. \ref{fig:Single Smarticle Locomotion as a function of gait amplitude})
Despite such restrictions, previous studies on collections of smarticles revealed diverse behaviors resulting solely from collisions. These included expansion, contraction and diffusion of  unconfined ``clouds" of smarticles and directed locomotion of confined smarticles (a phototaxing ``supersmarticle" \cite{savoie2019robot, savoie2018phototactic}); further these robots were used to test a new principle of active collectives, the spontaneous selection of ``low-rattling" states \cite{chvykov2021low}. 

Here we study in laboratory experiment and numerical simulation unconfined collections of smarticles and report the discovery of \emph{gliders}---dynamically bound pairs of smarticles which form spontaneously from a cloud and travel for long distances, interacting solely via locally repulsive rigid body collisions. These gliders are reminiscent of `Gliders' in Conway’s CA Game of Life \cite{gardner1970mathematical}, which are patterns that move via periodic changes to their configuration. Although an individual smarticle has limited motility, smarticle gliders travel superdiffusively as a result of the collisions caused by the periodic shape-changing motions of the smarticles. Experimentally validated simulations give insight into glider formation and transport properties, revealing principles that could potentially guide the development of dynamically entangling collectives for self assembly and transport in shape-changing active matter.

\emph{Experimental apparatus.}
The smarticles studied have a mass of \emph{m} = 34.8 $\pm$ 0.6g, and are composed of three links: two side arms of length \emph{L} = 5.4 cm and thickness of 0.3 cm, and a central link (body) of length \emph{W} = 5.0 cm and thickness of \emph{D} = 2.2 cm. The smarticles were placed on a $60 \times 60\, \mathrm{cm}^2$ aluminium plate, levelled to $\leq 0.1 ^{\circ}$. The gait, i.e., the sequence of shape-changing motions, of each smarticle, depicted in Fig.~\ref{fig:Discovery}A,B, was inspired by the dynamics of Purcell’s three-link swimmer \cite{purcell1977life,hatton2013geometric}. The periodic shape-changing moves are achieved by actuating the revolute joints connecting the central link to the side arms with a programmable servo motor with a gait period of 1.6\  sec. The side arms were rotated at the maximal motor speed until reaching the target arm rotation amplitude angle. 
Since the moving arms of the smarticle rest above the central link, they rotate without interacting with the ground, and the inertial impulses they impart on the central link typically induce negligible motion. The friction between the smarticle and the underlying surface limited the motion induced by the arm rotations to about 0.0015 $W$ (0.0075 cm) per cycle, where $W$ is the length of the central link. Therefore, individual smarticles do not move significant distances on their own.

\emph{Observation of Gliders.}
A collection of 7 smarticles was initialized in a densely packed configuration, where all smarticles began their shape-changing motions at the same time and in phase, as shown in Fig.~\ref{fig:Discovery}C. We hypothesized that the smarticles would push each other away via collisions thus expanding to a relaxed state where no smarticles are in contact with each other. Although previous studies of cloud dynamics \cite{savoie2019robot} had hinted at geometry-induced cohesion via fluctuations in the area fraction. It was frequently observed (77 out of 151 total trials) that a pair of smarticles would move together, as shown in Fig.~\ref{fig:Discovery}D (Supplementary Movie 1), for $\geq 100$ gait periods, traveling an average distance of $2.9 \pm 0.2$ $W$. 
The formation, stability and locomotion of these \emph{gliders} was remarkable given that the smarticles only interacted via collisions which are locally repulsive. The discrete and highly nonlinear nature of their collisions made it challenging to analyze them theoretically. In order to systematically study the gliders, a simulation was developed based on the open source physics engine DART \cite{Lee2018DART} (see SM for details), and calibrated against experiment (see Fig. \ref{fig:Calibration of Sim}). In the following sections we discuss the anatomy, robustness, binding mechanism, and locomotion of the gliders.

 \begin{figure}[ht!]
   \centering
   \includegraphics[width=1.0\columnwidth]{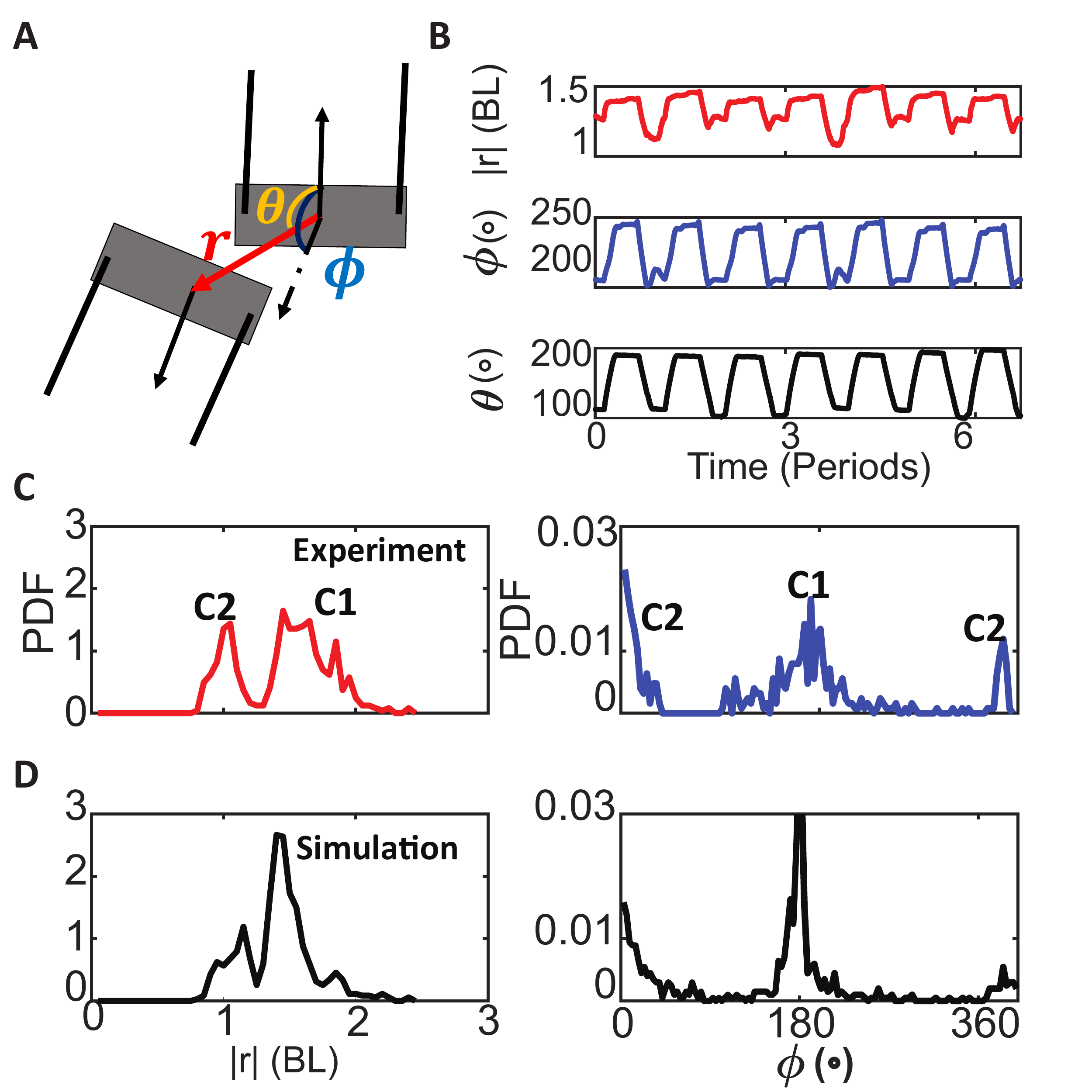}
   \caption{\textbf{Glider configurations.} 
   (A) Glider described by a pair of smarticles with relative position in polar coordinates ($r, \theta$), and relative body orientation ($\phi$) calculated by taking the angle between the two normal vectors.
   (B) Relative coordinates observed for a glider over 7 gait periods.
   (C-D) In both experiment and simulation, distributions of relative coordinates observed each gait period display two distinct clusters, C1 and C2.}
   \label{fig:GliderModes}
   \end{figure}

\emph{Glider Configurations.}
 Several (N = 151) trials of experiments and simulations revealed that the gliders existed in two distinct configurations. These configurations bear resemblance to the anti-phase and in-phase modes observed in Huygens' clocks \cite{bennett2002huygens} and oscillons \cite{umbanhowar1996localized} in vibrated granular media. We refer to them as :(i) ``$C1$" (anti-phase), in which the two smarticles are almost anti-aligned one with the other, their orientations differing by approximately $180$ degrees and (ii) ``$C2$" (in-phase) in which the two smarticles are oriented in approximately the same direction. Snapshots of glider configurations are depicted in Fig.~\ref{fig:Discovery}E. To characterize these distinct configurations, we measured their inter-smarticle distance $\vert \vec{r} \vert$, the polar angle $\theta$ and their relative orientation $\phi$, defined in Fig.~\ref{fig:GliderModes}A.
 {The time evolution of the relative coordinates $\vert r \vert$, $\theta$ and $\phi$ over 7 periods for a glider trajectory (Fig.~\ref{fig:GliderModes}B) showed consistent behavior over many periods, indicative of a limit cycle \cite{eldering_role_2016}}. The distribution of the relative coordinates of gliders observed in experiments and simulation showed distinct peaks in their histograms (Fig.~\ref{fig:GliderModes}C-D), corresponding to the distinct glider configurations $C1$ and $C2$ (Supplementary Movie 2). Further, the two glider configurations possessed distinct lifetimes (Fig.~\ref{fig:Lifetimes and contour lengths}), with  $C1$ gliders persisting longer than $C2$ gliders. Trials were conducted for 300 gait periods, and in some cases a $C1$ glider was still bound and locomoting when the trial was stopped. The lack of perfect synchronization in the gaits of the smarticle pair forming the glider in experiments lead to a higher variance in the distribution of the relative coordinates $r, \phi$ as compared to simulations (see Fig.~\ref{fig:GliderModes}C-D).

 \begin{figure}[ht!]
 \centering
  \includegraphics[width=1.0\columnwidth]{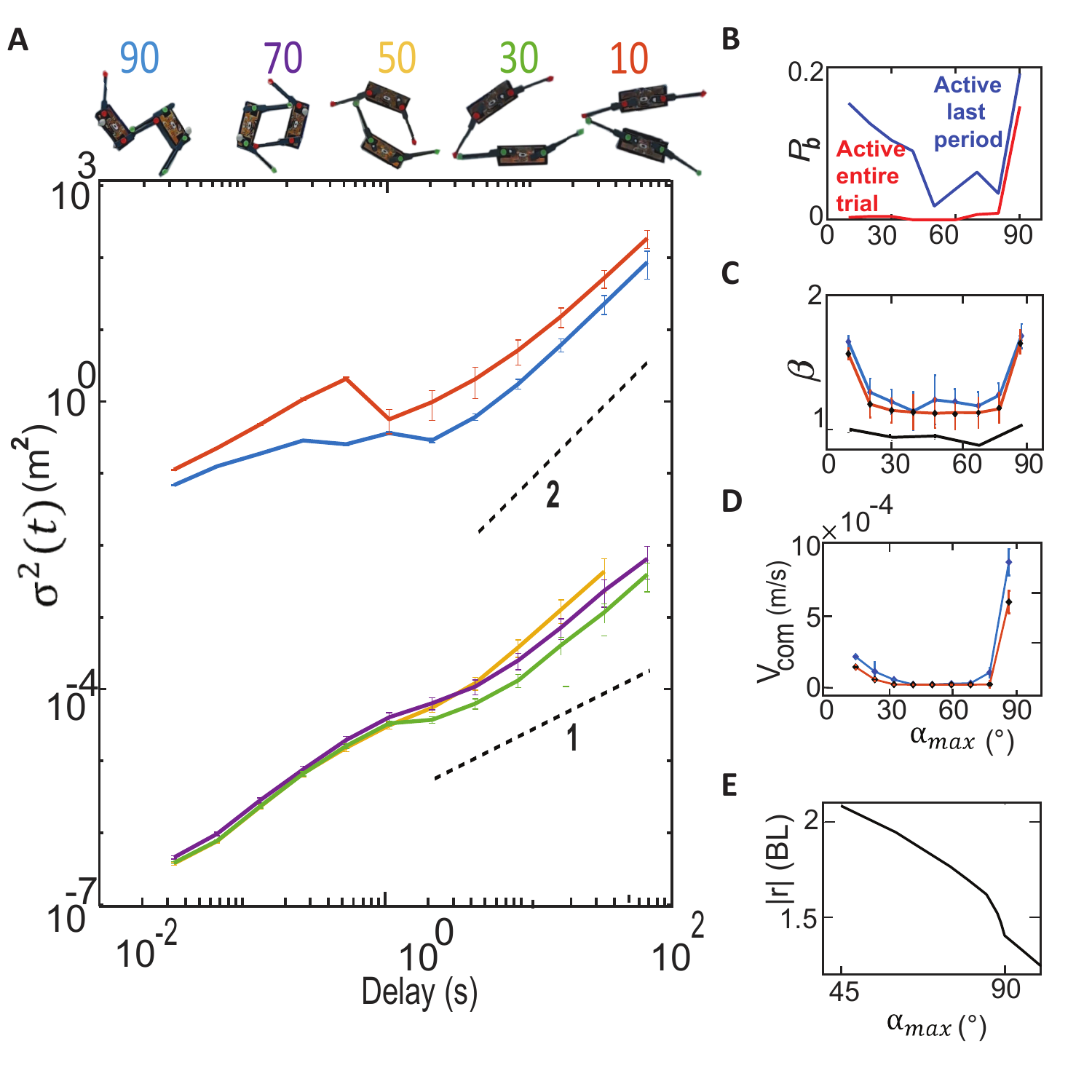}
  \caption{\textbf{Bound pair displacement dynamics}. (A) MSD $ \langle \sigma^2(t) \rangle   $   vs time-delay for 5 arm amplitude values with snapshots of the bound pairs shown besides the plot.  (B) Probability to bind \emph{$P_b$} vs. arm amplitude, for random initial configurations; blue curve represents the bound pairs found active at the end of the trial, red curve represents pairs that remained active throughout the trial.
  (C) MSD Exponents ($\beta$) vs. $\alpha_{max}$ for bound pairs, from msd slopes fit at long time delays. Blue and orange curves show experiment and simulation data, respectively; Bars represent standard-deviation. The black curve is the exponent calculated for a time symmetric gait (S1 to S3 to S1, see Fig. 1A) in simulations.
  (D) Bound pair center-of-mass speed ($V_{com}$) vs.$\alpha_{max}$ .
  (E) Steady state separation $|r|$ vs. $\alpha_{max} > 45^{\circ}$, from simulations of smarticles started off in bound states. }
  \label{fig:BoundPairTransport}
\end{figure}

\emph{ Glider robustness and transport.}
To test the robustness of bound-pair states and their subsequent locomotion, we systematically varied the arm amplitude, i.e. the maximum rotation angle $ \alpha_{max} $ of the smarticle arm, while keeping the smarticle arm rotation speed fixed and studied the persistence of bound pairs for shapes with varying degrees of concavity \cite{rosenfeld1985measuring}. We performed experiments and numerical simulations, fixing one smarticle and initializing the other in a polar grid of constant radius with 1000 points in simulations and 21 points in experiments (see Fig.~\ref{fig:Basin Scan Pictorial}) for a range of $ \alpha_{max} $. We observed a rich repertoire of bound states i.e., where smarticles made contact at least once during their gait, for all amplitudes, suggesting that these bound states robustly form for a variety of $ \alpha_{max} $. 

The mean-square displacement (MSD) vs time plot of all glider trajectories (surviving greater than 30 cycles) for 5 different arm amplitudes is shown in Fig \ref{fig:BoundPairTransport}A. The bound pairs for $90^{\circ}$ and $10^{\circ}$ move nearly ballistically with the  $ \langle {\sigma } ^ 2(t) \rangle  \propto t ^ {\beta} $  (where MSD $ \langle {\sigma } ^ 2(t) \rangle = \langle \vec{x}^2(t) \rangle -{\langle \vec{x}(t) \rangle}^2  $ with exponent $\beta \approx 1.8$), 
while for the intermediate values of arm amplitude the motion is almost diffusive. We observed that gliders studied in experiments have a higher $ \beta $ and $ V_{com} $ gliding speed (speed of the com of the 2 particle bound excitation) than those studied in simulations as shown in Fig.~\ref{fig:BoundPairTransport}. We posit that the noise in experiments from the motors, the collisions and the friction with the floor enhances the transport of the bound pairs with intermediate arm amplitudes ($40^{\circ} - 60^{\circ}$), but the pairs bind and unbind frequently. This lack of stability for intermediate $ \alpha_{max} $ correlates with a decrease in binding probability (Fig.~\ref{fig:BoundPairTransport}B). The different bound states in experiments and simulations for the three different amplitude regimes are shown in Supplementary Movie 3.

Bound pair speed and transport decreased sharply as  $ \alpha_{max} $ decreased from $90^{\circ}$, but surprisingly increased again in the low amplitude regime (see $10^{\circ} - 20^{\circ}$ range in Fig.~\ref{fig:BoundPairTransport}C-D). We attribute this anomaly to inertial effects  which enhanced the drift of a single smarticle as shown in Fig.~\ref{fig:Single Smarticle Locomotion as a function of gait amplitude}. Since the smarticle arms were driven at a constant motor speed, for low rotation angles the time of rotation of a smarticle arm became comparable to the time of overlap between the motions of two arms as shown in
Fig.~\ref{fig: Gait amplitude}. This led to a significant increase in the motility of an individual smarticle. Consequently gliders at low  $ \alpha_{max} $ also showed an improvement in their gliding speed as shown in Fig.~\ref{fig:BoundPairTransport}(C-D). 

To probe how arm amplitude affects glider locomotion, we performed simulations of pairs of smarticles in a bound formation, performing gaits with varying $\alpha_{max}$ for values greater than $45^{\circ}$. A clear trend appears, where the distance monotonically decreases with increased $\alpha_{max}$ Fig.~\ref{fig:BoundPairTransport}E,  with a rapid drop in distance around $85^{\circ}$. Coincidentally, this is also roughly the amplitude at which the $\beta $ and $ V_{com} $ shoot up (Fig. \ref{fig:BoundPairTransport}C-D). Simulations revealed that for arm amplitudes below $85^{\circ}$ the smarticles only interact with one arm, and above they experience an additional collision -- with the other arm, as depicted in the middle configuration of the attractive case in Fig. \ref{fig:BindingMechanism}C and Supplementary Movie 3. The additional collision seems to result in closer proximity between the smarticles, which leads to a third collision between an arm and the middle link. Since super-diffusive transport becomes effective above arm amplitude of $85^{\circ}$, we attribute most of the locomotion to these additional collisions.

\emph{Binding mechanism and factors responsible for gliding.}
Though two convex-shaped bodies interacting only via repulsive collisions must repel ( Fig. 4A), pairs of smarticles undergoing repulsive collisions can remain bound since they develop concave shapes during their shape-changing motions. This makes it possible for the net contact impulse during a collision of two smarticles to have an attractive effect $((\vec {I}_N + \vec {I}_F)\cdot\vec{r}>0) $ for atleast one collision during the gait, typically this is when the 2 red arms collide for the C1 glider (Fig.~\ref{fig:BindingMechanism}A) and results in the two smarticles coming closer. Friction between the smarticles and the base plate prevents coasting, and the periodic attractive contact forces during collisions ensure they remain bound over multiple gait periods.
Most of the stable glider configurations converge to the periodic deformation pattern characteristic of the attractor within a few periods; moreover, their ultimate fate -- binding or not -- could be determined from the configuration following the first gait period - which orients the robots such that they are either attracted or repelled  by the subsequent collisions.  We show the first 5 periods of 2 representative initial conditions stroboscopically at the same phase of the gait with the corresponding $|r|$ values (Fig.~\ref{fig:BindingMechanism}B) to illustrate how collisions in the same phase of the gait cycle could produce binding vs. repulsion depending on how the smarticles started off relative to each other.

To identify the conditions leading to binding, we performed simulations for 5000 uniformly randomly sampled configurations in the  $(\vert r \vert,\theta, \phi)$ space (Fig.~\ref{fig:BindingMechanism}C). Once initialized, we tracked the lifetimes of bound pairs for 75 gait periods and separated the relative configuration space into regions of attraction vs. repulsion based on the fate of the corresponding smarticle duo. The glider configurations $C1$ and $C2$ correspond to attractors in these selected regions of the configuration space. Fig.~\ref{fig:BindingMechanism}D shows the structure of the attractor in chosen coordinates, with corresponding projections on the axes. 
\begin{figure}[ht!]
 \centering
  \includegraphics[width=0.8\columnwidth]{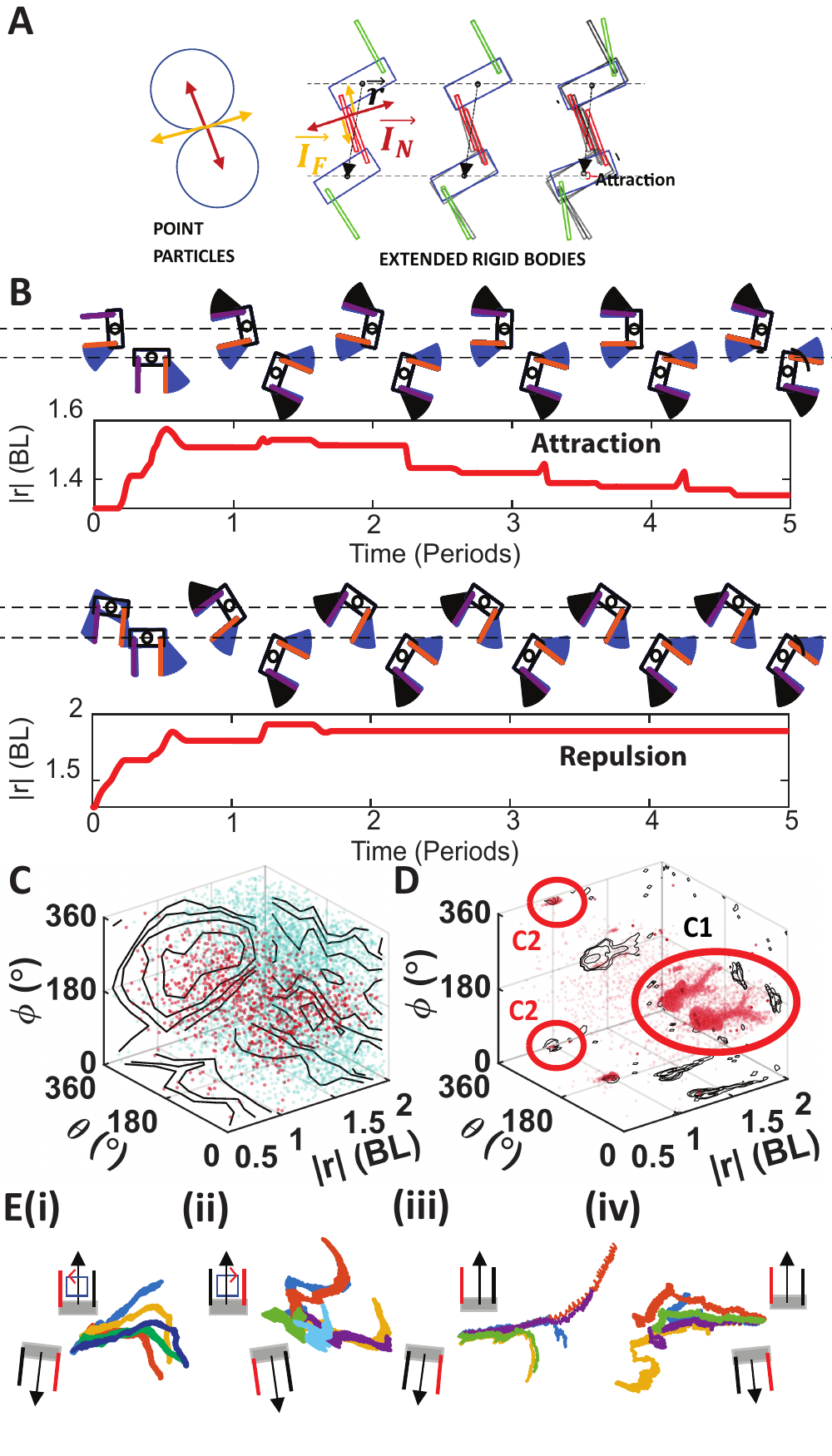}
  \caption{\textbf{Interactions leading to binding and locomotion.}
  (A) Schematic showing how locally repulsive collisions can produce attraction in extended rigid bodies undergoing shape change contrasted with point particles. Color represents current frame while grey indicates past frame.
  (B) Stroboscopic representation of 2 initial conditions leading to attraction and repulsion respectively for the first 5 periods with the corresponding $|r| $. Blue cones represent the trajectory of the arms 20 frames into the future, while black cones represent the same 20 frames in the past.
  (C) Simulations of smarticle pairs initialized in random configurations (all points), of which many led to stable gliders (red points).
  (D) At steady state the configurations corresponding to the stable gliders show up as distinct attractors for the C1 and C2 type gliders.
  (E) Role of  relative position and orientation E(i,ii) and gait chirality E(iii,iv) (asymmetries) on glider locomotion.}
  \label{fig:BindingMechanism}
\end{figure}

The basin of attraction for the $C1$ glider was larger than that for the for the $C2$ type glider. As a result, anti-aligned smarticle pairs were likely to form a $C1$ type glider if initialized sufficiently close together. 
Fig.~\ref{fig:Basin Scan Pictorial} (A) shows the schematic for the initial condition scan and Supplementary Movie 4 shows the flow of the initial configurations onto the attractors that form the gliders. Though glider formation is highly likely, the boundary between attractive and repulsive initial conditions depends on relative angle and position, at a given phase, in a non-trivial way. See Fig.~\ref{fig:Basin Scan Pictorial} (B-C) for details.


Glider trajectories are governed by gait chirality and configurational asymmetry. The chirality of the performed gait results in a consistent pattern of arm to arm collisions where, starting from a non-overlapping configuration (``facing away'', as in Fig. \ref{fig:BindingMechanism}E(i)), the first arm to collide with the opposing arm is always the same one (shown in red). When smarticles are programmed to perform a gait with the opposite chirality, they form glider pairs that engage first with the opposite arm (Fig. \ref{fig:BindingMechanism}E(ii)). Gliders of a given chirality are a mirror image of gliders of the opposite chirality. It is therefore expected that the resulting trajectories will diverge in opposing directions, and as seen in multiple trials, they do (Fig. \ref{fig:BindingMechanism}E(i, ii)). In addition, the chirality of each gait appears to manifest in the chirality of the resulting trajectories. Imposing a time reversible gait makes the system lose persistent locomotion and the impending motion of the bound state turns diffusive. We performed experiments and simulations with time symmetric gaits (diagonal in the configuration space, S1 to S3 to S1 in Fig. \ref{fig:Discovery}) for varying arm amplitude, as mentioned above, and observed that the value of $\beta$ was close to one (black curve in Fig. \ref{fig:BoundPairTransport} C). See Supplementary Movie 5 for an example of a time-symmetric gait.
 
Although glider pairs appear symmetric under $180^{\circ}$ rotation (and half gait period shift for C2). Such symmetry should not result in a directional bias, yet gliders exhibit persistent, nearly ballistic, locomotion, in which one smarticle consistently appears somewhat ahead of the other. Upon observation it is revealed that relative orientation between the smarticles deviates from $180^{\circ}$ (Fig. \ref{fig:BindingMechanism}E(iii)), in a consistent manner. The direction of this deviation from (anti-)alignment controls the direction of the trajectory, as evident from multiple trials (Fig. \ref{fig:BindingMechanism}E(iii, iv)). This is expected, since the configuration in Fig. \ref{fig:BindingMechanism}E(iv) can be obtained by an almost $180^{\circ}$ rotation of the configuration in Fig. \ref{fig:BindingMechanism}E(iii), this effect is also evident in the structure of the C1 attractor as the two clusters which have an apparent reflection symmetry (Fig. \ref{fig:BindingMechanism}D).

\emph{Conclusion.}
We studied ensembles of shape-changing robots (smarticles) in experiments and simulations, and reported the novel phenomenon of \emph{gliders}---spontaneously formed dynamically bound pairs of smarticles that travel together with almost ballistic locomotion, over multiple body lengths, enduring hundreds of rigid-body collisions between them. We characterized gliders by the relative spacing and orientation between their smarticles and showed that there is a vast number of initial configurations that bind and locomote. We attribute the nearly ballistic nature of glider motion to the breaking of time-reversal symmetry by their gait, as well as the spontaneously broken rotational symmetry of their configuration. Further, we showed that gliders exist and travel superdiffusively for a wide range of arm amplitudes, a parameter which essentially controls the degree of concavity manifest by smarticles, as they changed their shape. Surprisingly these physical gliders also display  some other unexpected similarities with their CA counterparts, like assembling into long structures upon collision with other smarticles in their vicinity (Supplementary Movie 6).  Our study suggests that gliders are likely to arise in a wide variety of active matter systems where shape-changing particles take concave shapes with time irreversible gaits in a sufficiently damped environment and adds further evidence of rich dynamics in active collisions.

\emph{Acknowledgments:} We thank Will Savoie for discovery of gliders, Pavel Chvykov and Zachary Jackson for stimulating discussions on smarticles, Yunbo Zhang for help with learning DART, and all members of the lab for support on miscellaneous fronts. Support provided by ARO MURI grant W911NF-13-1-034, School of Physics Georgia Institute of Technology, QBIOS PhD program and a Dunn Family Professorship (D.I.G).


\title{Gliders in shape-changing active matter \\
\large Supplementary Information\\
}

\author{Akash Vardhan}
\affiliation{%
 School of Physics, Georgia Institute of Technology, Atlanta, Georgia 30332, USA}%
\author{Ram Avinery}%
\affiliation{%
 School of Physics, Georgia Institute of Technology, Atlanta, Georgia 30332, USA}%
 \author{Hridesh Kedia}%
\affiliation{%
 School of Physics, Georgia Institute of Technology, Atlanta, Georgia 30332, USA}%
 \author{Shengkai Li}%
\affiliation{%
 School of Physics, Georgia Institute of Technology, Atlanta, Georgia 30332, USA}%
 
\author{Kurt Wiesenfeld}
\affiliation{%
 School of Physics, Georgia Institute of Technology, Atlanta, Georgia 30332, USA}%
 
 \author{Daniel I. Goldman}%
\affiliation{%
 School of Physics, Georgia Institute of Technology, Atlanta, Georgia 30332, USA}%

\date{\today}

\maketitle


\section{SUPPLEMENTARY INFORMATION}

\begin{figure}[ht!]
  \centering
  \includegraphics[width=1.00\columnwidth]{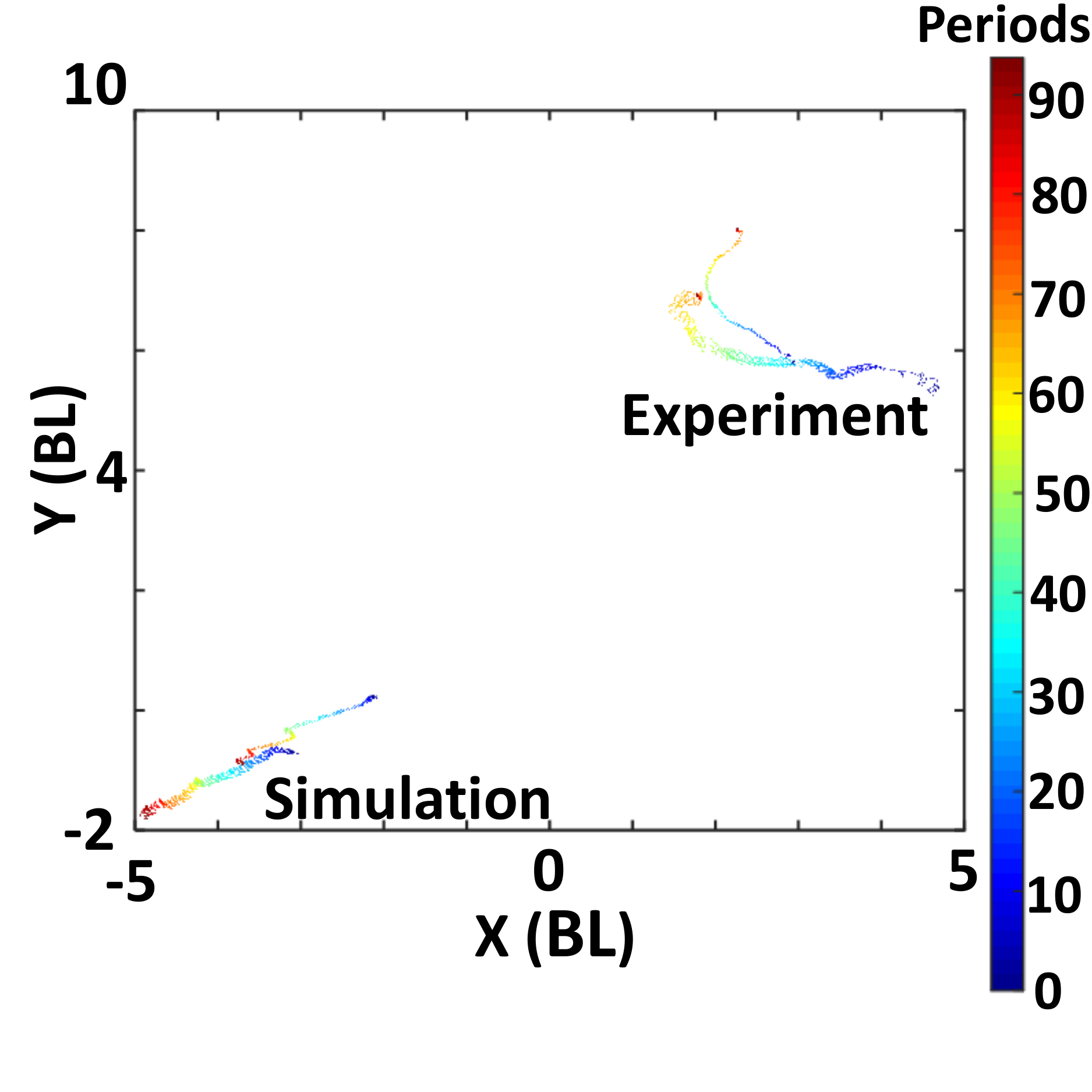}
  \caption{ Representative trajectories from an experimental and simulated trials over  90 gait periods. }
  \label{fig:Calibration of Sim}
\end{figure}

\emph{Comparison between simulation and experimental tracks.}
A simulation of the smarticles  was created using the open source physics engine Dart to seek understanding of the phenomena. The rigid body equations of motion are formulated as an implicit time-stepping, velocity based LCP (Linear complimentarity problem) which are then solved using standard constraint solvers like the Lemke and Dantzig methods, after \cite{stewart2000implicit} . The time step used in the simulations was 1e-4 seconds which guaranteed the best convergence of trajectories between simulations initialized from a set of initial conditions. Fig.~\ref{fig:Calibration of Sim} shows the tracks of the body centers of the 2 smarticles for 90 periods in experiment and simulation. Experimental tracks tend to curve more which can be attributed to the various sources of noise in the real world, in our simulations we did not add any external noise (over the accumulated numerical error), since we wanted to understand the phenomena in a cleaner more pristine setting, the tracks in the simulation have much lower direction modulation consequently, which also shows up as tighter peaks in the relative coordinate PDF's as shown in ~Fig. \ref{fig:GliderModes} C.

\begin{figure}[ht!]
  \centering
  \includegraphics[width=1.00\columnwidth]{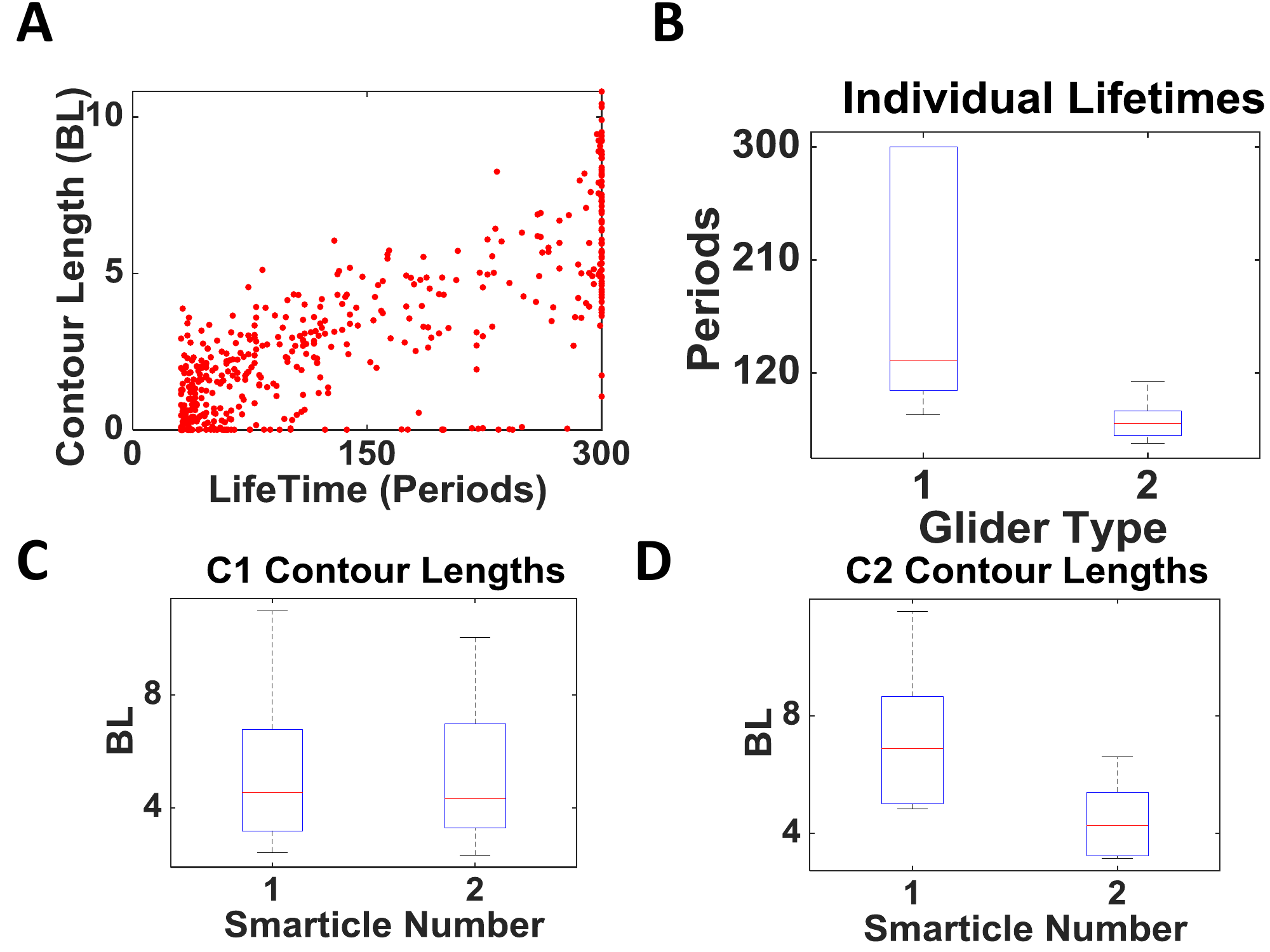}
  \caption{ Trajectory contour length and lifetime for emergent gliders. (A) Contour length vs. contour time for all bound pairs emerging from the relaxation experiments (see Fig. 1), for experimental trials recorded over 300 gait periods. 
  (B) Lifetimes of the glider configurations in the relaxation experiments.
  (C-D) Contour lengths of C1 (C) and C2 (D) type gliders after initializing them from known initial configurations, over 21 trials. }
  \label{fig:Lifetimes and contour lengths}
\end{figure}

\emph{Lifetimes and contour lengths of emergent bound pairs.}
It usually took the closed packed configurations about 30 periods to reach steady state. So, to characterize the lifetime of our bound pairs we decided to plot the lifetimes vs. displacements (contour length) of all 2 smarticle bound pairs Fig. \ref{fig:Lifetimes and contour lengths} A and observed a linear correlation between the two suggesting that the longest surviving bound pairs would also travel the largest distance. We then characterized the lifetime of individual glider configurations and observed that the C1 type gliders were more stable and lived longer than the C2 counter parts Fig. \ref{fig:Lifetimes and contour lengths} B. To measure the displacements of individual glider types we initialized the smarticles from known configurations, and observed that for the C1 glider the 2 smarticles would travel roughly the same distance  Fig. \ref{fig:Lifetimes and contour lengths} C,  while in the C2 type glider there would constantly be one smarticle which would end up travelling more than the other as they would curve as seen in Fig. \ref{fig:Lifetimes and contour lengths} D and Supplementary movie 2. The C2 configuration was susceptible to tiny phase differences in the arms of the 2 smarticles, and required near perfect synchronization, as a result the 2 smarticles would break apart from the C2 configuration causing them to have a lower lifetime than their C1 counterparts, as shown in Fig. \ref{fig:Lifetimes and contour lengths}B and supplementary movie 2.

\begin{figure}[ht!]
  \centering
  \includegraphics[width=1.00\columnwidth]{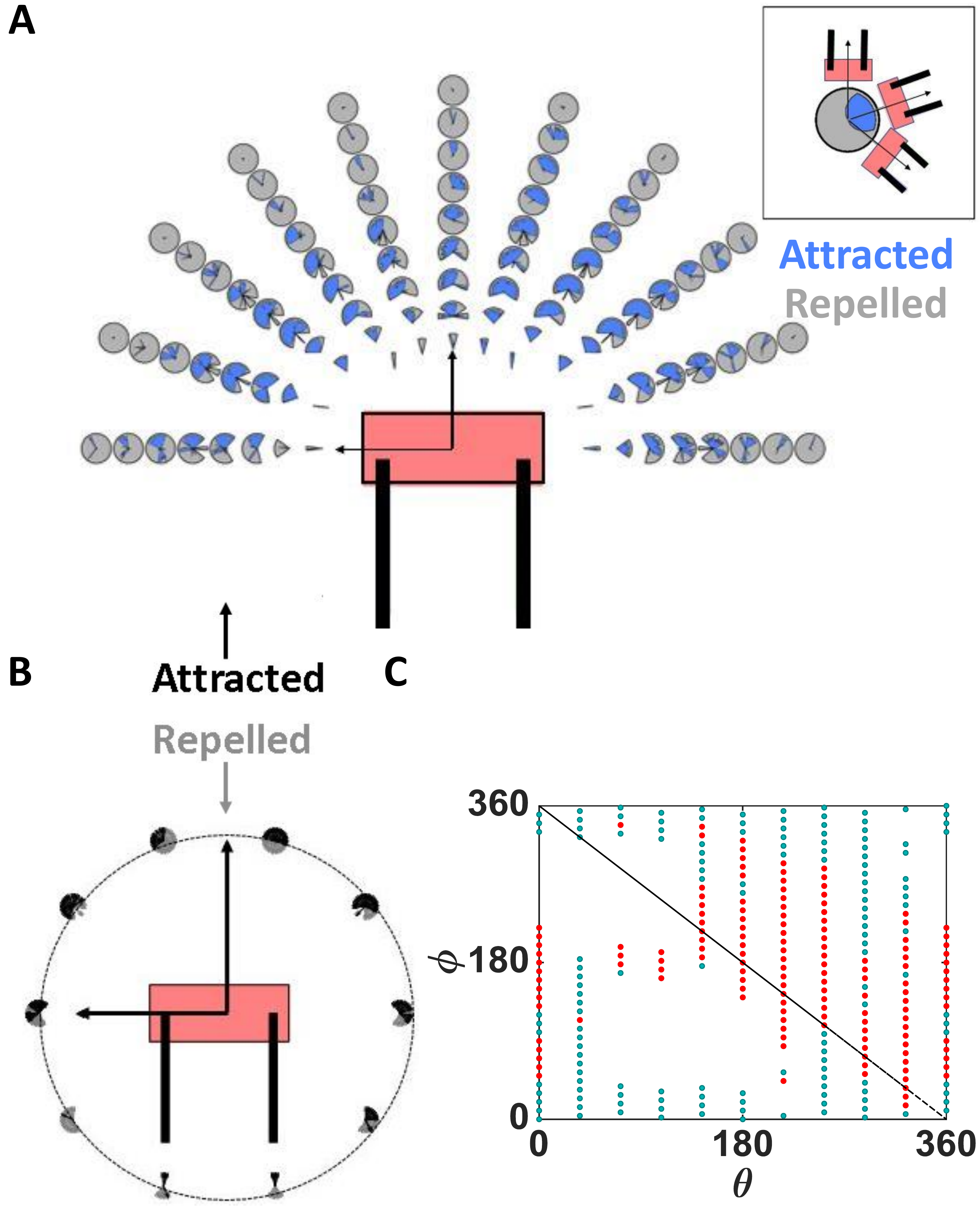}
  \caption{ Configurations leading to attraction. (A) Smarticle pair configurations were sampled on a polar grid of positions on the the upper half circle around  a fixed smarticle. At each position, the orientation of the placed smarticle was uniformly sampled. The polar plot at each position represents valid (nonintersecting, filled - gray or blue) or invalid (white) configurations, where blue (gray) indicates attraction (repulsion). (B) Smarticle pair configurations were also sampled at a constant radius, in all positional angles and uniformly over orientation. Polar plots are same as in A.
  (C).\emph{Angular coordinate phase diagram.} Initial configurations in B that led to attraction are depicted as red, while those that led to repulsion are depicted as cyan points, with white being the physically invalid configurations over the positional ($\theta$) and orientation ($\phi$) angles.  A dashed line of $\phi = (360 -  \theta)^{\circ} $ was added as a visual aid.} 
  \label{fig:Basin Scan Pictorial}
\end{figure}

\emph{Pictorial representation of basin scan.}
To probe the binding mechanism and the robustness of gliders we initialized smarticles, by fixing one and varying the position of the second one around it in polar coordinates and also the absolute orientation and separated the configuration space into regions of attraction vs. repulsion. This provided us with the control and resolution to zoom into the events within a gait period and understand how the same collisions bring about the 2 effects due to the geometrical configurations they start off with. Fig \ref{fig:Basin Scan Pictorial} (A-B) shows how we set up the scan, and the corresponding configurations which got attracted vs. repelled. To demonstrate positional angle and orientation dependence, we show scan results in Fig \ref{fig:Basin Scan Pictorial} (B-C), for a fixed radius of 1.3 $W$. Fig \ref{fig:Basin Scan Pictorial} (C) shows a phase diagram in the initial angular coordinate space, with cyan corresponding to the repelled configurations, red corresponding to configurations that ended up forming bound pairs and white being the invalid configurations, of the physical configurations sampled in Fig \ref{fig:Basin Scan Pictorial} B. The attracted configurations seem to be almost symmetrically distributed about the normal vector of the fixed smarticle, and this showed up as the symmetry in the attracted configurations about ($ \theta \approx 270^{\circ} )$ .

\begin{figure}[ht!]
  \centering
  \includegraphics[width=1.00\columnwidth]{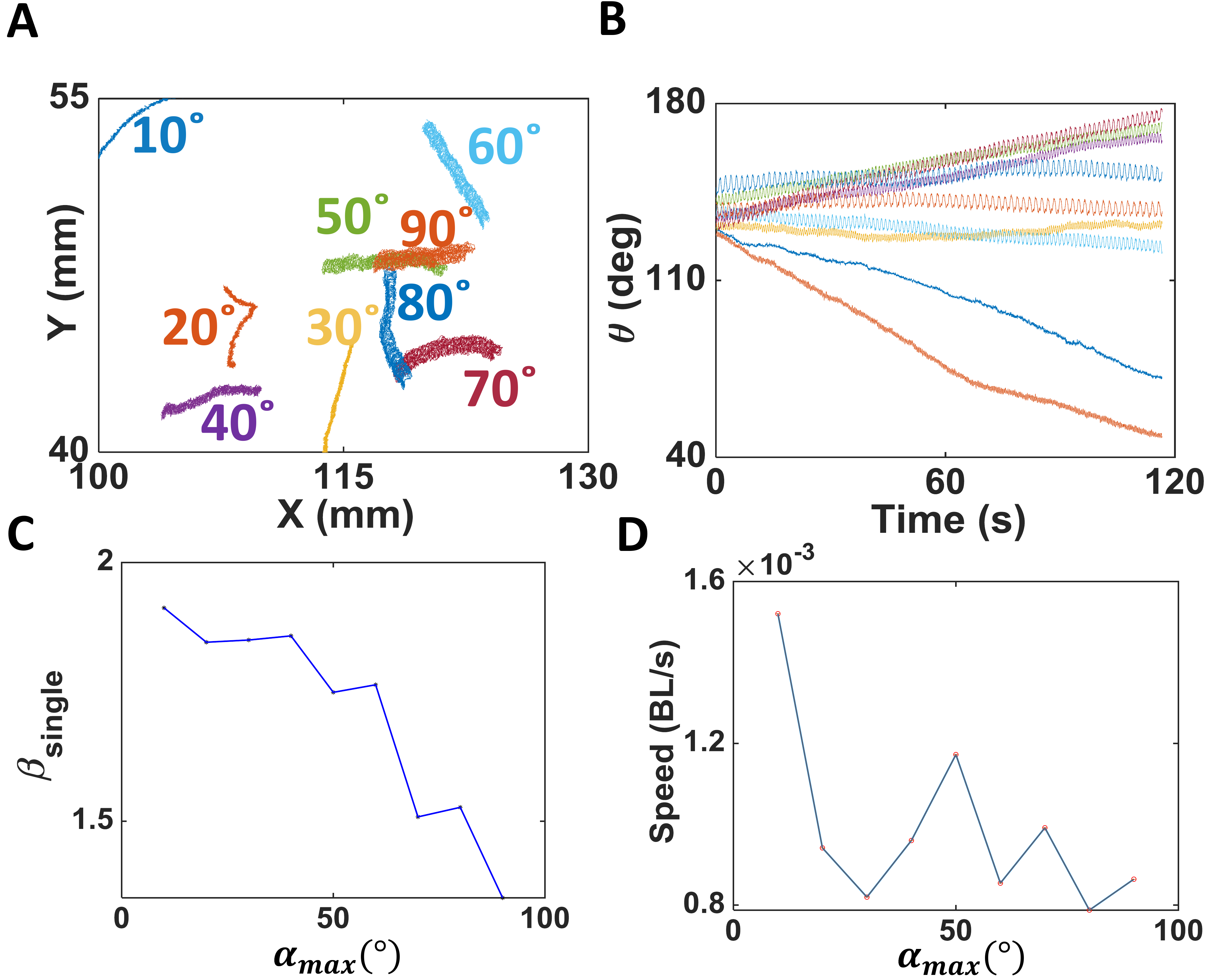}
  \caption{Motion of individual smarticles as a function of gait amplitude. (A) Tracks of the body center of the smarticle for different gait amplitudes, over 75 gait periods. (B) Time series for the angular coordinate of the smarticle. (C) $\beta_{single}$ (MSD exponent) of a single smarticle vs. arm amplitude. (D) Speed of the body center vs. arm amplitude.}
  \label{fig:Single Smarticle Locomotion as a function of gait amplitude}
\end{figure}

\emph{Single Smarticle Locomotion vs Amplitude.}
The anomalous transport of the bound pairs that were formed at lower arm opening angles was attributed mostly due to the enhanced drift of a single smarticle at lower amplitudes. We wanted to characterize the behavior of a single robot as a function of the amplitude. Fig \ref{fig:Single Smarticle Locomotion as a function of gait amplitude} (A-B) depict the body center tracks and the angular drift of a single robot for the different gait amplitudes. The motion of a single smarticle is super-diffusive at higher and intermediate amplitudes and approaches the ballistic regime for lower values of the amplitude and it caused an increased drift speed, as shown in Fig \ref{fig:Single Smarticle Locomotion as a function of gait amplitude} (C-D). 

\begin{figure}[ht!]
  \centering
  \includegraphics[width=1.00\columnwidth]{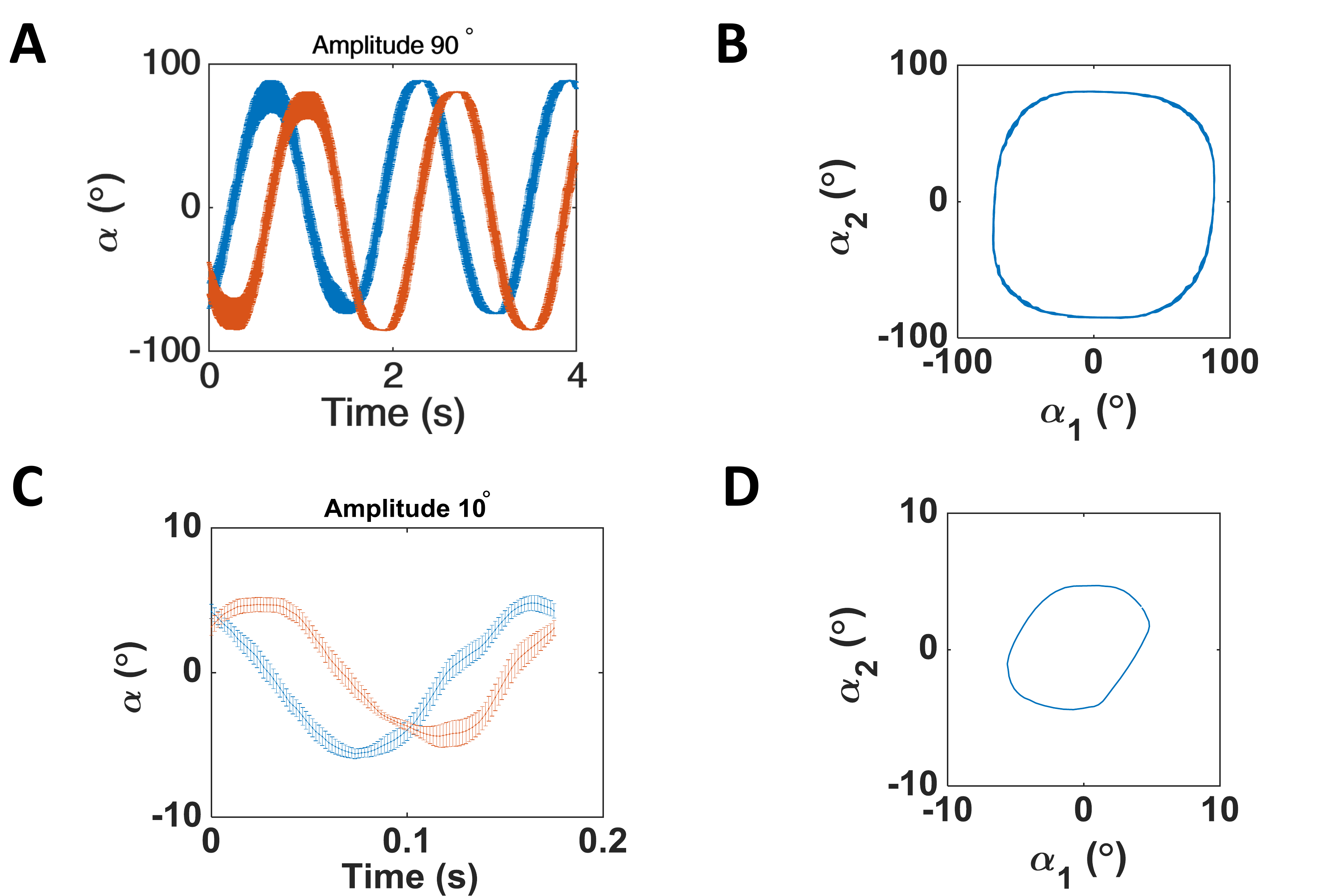}
  \caption{ Experimentally tracked arm angles for 90 and 10 degree amplitude. 
  (A,C) Time-series of the first arm angle ($\alpha_1$) over 3 gait periods for 90$^{\circ}$ (A) and a single period for 10$^{\circ}$ (C) degree arm amplitudes. Blue and orange curves represent arms 1 and 2 respectively.
  (B,D) Gait represented in arm-angle configuration space ($\alpha_1, \alpha_2$), averaged over 10 periods, for 90 (B) and 10 (D) degree arm amplitudes. }
  
  \label{fig: Gait amplitude}
\end{figure}

\emph{Experimental track of arm angles of single smarticle.}
We hypothesized that this anomaly was due to inertial effects from the overlap in the motions of the 2 arms, since the gait was executed at the maximal motor speed for all arm amplitudes and we adjusted the dwelling time at the corners of the square to ensure the time period scaled commensurately with the amplitude. At lower amplitudes, the value of this delay was so small that the arms wouldn't get to the prescribed angle, and the other arm would move before the first arm completed its motion. This caused a drift in the center of the middle link. Fig. \ref{fig: Gait amplitude} (A,C) shows experimentally tracked arm angles of a single smarticle for 90 and 10 degrees, and when plotted in the shape configuration space \ref{fig: Gait amplitude} (B,D), it becomes clear that in the low amplitude regime the arms don't finish the whole prescribed range of motion.

\bibliography{Glider_Ref}
\end{document}